# An Agent-Based Simulation of Regularity-Driven Student Attrition: How Institutional Time-to-Live Constraints Create a Dropout Trap in Higher Education

Hugo Roger Paz
PhD Professor and Researcher Faculty of Exact Sciences and Technology National University of Tucumán
Email: hpaz@herrera.unt.edu.ar
ORCID: https://orcid.org/0000-0003-1237-7983

**Abstract**

High dropout rates in engineering programmes are conventionally attributed to student deficits—lack of academic preparation or motivation. However, this view neglects the causal role of "normative friction": the complex system of administrative rules, exam validity windows, and prerequisite chains that constrain student progression. This paper introduces "The Regularity Trap," a phenomenon where rigid assessment timelines decouple learning from accreditation. We operationalize the CAPIRE framework into a calibrated Agent-Based Model (ABM) simulating 1,343 student trajectories across a 42-course Civil Engineering curriculum. The model integrates empirical course parameters and thirteen psycho-academic archetypes derived from a 15-year longitudinal dataset. By formalizing the "Regularity Regime" as a decaying validity function, we isolate the effect of administrative time limits on attrition. Results reveal that 86.4% of observed dropouts are driven by normative mechanisms (expiry cascades) rather than purely academic failure (5.3%). While the overall dropout rate stabilized at 32.4%, vulnerability was highly heterogeneous: archetypes with myopic planning horizons faced attrition rates up to 49.0%, compared to 13.2% for strategic agents, despite comparable academic ability. These findings challenge the neutrality of administrative structures, suggesting that rigid validity windows act as an invisible filter that disproportionately penalizes students with lower self-regulatory capital.

.

# 1. Introduction

## 1.1 The Global Challenge of University Dropout

University dropout represents one of the most pressing challenges confronting contemporary higher education systems, with profound implications for individual life trajectories, institutional sustainability, and societal equity. Across OECD nations, approximately 30–40% of students who commence tertiary education fail to complete their degrees [OECD, 2019), whilst in Latin America, attrition rates frequently exceed 50% in public universities [UNESCO-IESALC, 2020). These statistics mask considerable heterogeneity: dropout patterns vary systematically by socioeconomic status, ethnicity, gender, and prior academic preparation, suggesting that attrition is not merely an individual failure but a structural phenomenon shaped by institutional design and policy choices (Tinto, 2017; Braxton & Hirschy, 2020).

The consequences of university dropout extend beyond foregone credentials. Students who withdraw prematurely often experience diminished lifetime earnings, reduced career mobility, and psychological distress associated with perceived failure [Bernardo et al., 2021). For institutions, high attrition rates strain resources, erode reputation, and complicate strategic planning. At the societal level, elevated dropout concentrates human capital losses amongst disadvantaged groups, perpetuating intergenerational inequality and limiting economic dynamism(Bound et al., 2020). Consequently, understanding the mechanisms that drive student attrition—and identifying modifiable policy levers to mitigate these forces—constitutes a research priority with substantial real-world impact.

## 1.2 Limitations of Prevailing Explanatory Models

Dominant theoretical frameworks for understanding university dropout have historically emphasised individual-level factors: academic under-preparation, motivational deficits, inadequate study skills, or insufficient social integration (Tinto, 1975; Bean & Metzner, 1985; Cabrera et al., 1993). Whilst these models have generated valuable insights, they suffer from three key limitations that constrain both theoretical understanding and policy relevance.

First, **individual-centric models risk attributional bias**, locating the "problem" of dropout within students rather than examining how institutional structures, policies, and resource allocation decisions shape attrition patterns [Bensimon, 2007; Witham & Bensimon, 2012). This framing obscures the extent to which dropout may reflect rational responses to systemic barriers—such as inflexible course scheduling, burdensome administrative requirements, or financial aid discontinuities—that disproportionately affect students from marginalised backgrounds Goldrick-Rab, 2016).

Second, **prevailing models struggle to isolate causal mechanisms** amidst the multicollinearity inherent in observational educational data. Students who withdraw typically differ from persisters across numerous correlated dimensions (e.g., academic ability, socioeconomic status, motivation, institutional support access), making it difficult to disentangle whether dropout results from intrinsic student characteristics, environmental stressors, institutional policies, or complex interactions amongst these factors Schneider & Yin, 2011). Traditional regression-based approaches, whilst useful for identifying associations, cannot definitively establish causation in such high-dimensional, interdependent systems Pearl & Mackenzie, 2018).

Third, **existing frameworks inadequately theorise the temporal dynamics of attrition**. Dropout is not a discrete event but rather the culmination of a dynamic process unfolding over multiple academic cycles, during which students experience fluctuating motivation, accumulating stress, evolving perceptions of belonging, and shifting opportunity costs Davidson et al., 2015; Tight, 2020). Static cross-sectional analyses or even longitudinal regression models that treat time as a covariate fail to capture the feedback loops, threshold effects, and path-dependent trajectories that characterise student progression Hovdhaugen, 2015).

### 1.3 Institutional Friction as a Structural Determinant

Emerging scholarship has begun to foreground **institutional friction**—the cumulative burden of administrative, bureaucratic, and regulatory obstacles that students must navigate to maintain enrolment—as a critical yet under-examined driver of attrition Castleman & Page, 2016; Page & Scott-Clayton, 2016). Institutional friction manifests through myriad mechanisms: complex financial aid applications, opaque degree requirement tracking, rigid course prerequisites, inconvenient office hours for support services, and—central to this study—**time-to-live (TTL) expiry regulations** governing the duration of course regularisation.

In many higher education systems, particularly in continental Europe and Latin America, students must not only pass examinations but also maintain "regular" status in enrolled courses by satisfying attendance requirements, submitting assignments, or sitting examinations within prescribed windows. **Regularity expiry** occurs when a student fails to meet these conditions within a stipulated timeframe (e.g., two examination cycles), at which point the course is administratively removed from the student's active curriculum, and re-enrolment becomes necessary. Whilst ostensibly designed to encourage timely progression, regularity expiry introduces a **structural vulnerability**: students may lose access to courses not due to academic failure per se, but because life circumstances (employment demands, caregiving responsibilities, health issues, financial constraints) prevented them from meeting procedural deadlines.

Crucially, regularity expiry operates as a **ratchet mechanism**. Each expiry typically triggers administrative penalties (course removal, enrolment delays, fee obligations), psychological costs (perceived failure, diminished self-efficacy, eroded belonging), and logistical complications (schedule conflicts with re-offered courses, prerequisite bottlenecks). As expiries accumulate, students experience escalating stress, dwindling course options, and lengthening time-to-degree, creating a self-reinforcing cycle wherein initial setbacks progressively constrain future success Denning et al., 2019). We term this phenomenon the **"regularity trap"**: a structural condition in which students become ensnared by procedural constraints that are largely orthogonal to their academic capability.

**1.4 Research Objectives and Contributions**

Despite the theoretical salience of institutional friction mechanisms, empirical research has been hampered by two challenges. First, **observational data confound policy effects with selection effects**: students who experience multiple regularity expiries also tend to differ systematically in observable and unobservable characteristics, making causal inference difficult. Second, **experimental manipulation of institutional policies is ethically and logistically infeasible**: universities cannot randomly assign students to face differential TTL constraints merely to evaluate policy impacts.

**Agent-based modelling (ABM)** offers a methodological solution to these challenges. ABMs simulate populations of heterogeneous agents (in this case, students) operating within a defined environment (the university's curricular and regulatory structure) according to explicit behavioural rules Bonabeau, 2002; Railsback & Grimm, 2019). By parameterising agent characteristics and decision algorithms using empirical data, ABMs can isolate the causal effect of specific policy mechanisms—such as regularity expiry rules—whilst controlling for confounding heterogeneity. Moreover, ABMs naturally accommodate the temporal dynamics, feedback loops, and emergent phenomena that characterise educational trajectories Hamill & Gilbert, 2016).

This study deploys an agent-based simulation to address the following research questions:

**RQ1:** To what extent do time-to-live (TTL) expiry regulations drive student dropout independently of academic performance?

**RQ2:** How does the impact of regularity expiry vary across student archetypes characterised by differential planning horizons, academic ability, and psychological resilience?

**RQ3:** What is the temporal profile of dropout events, and does evidence exist for a "regularity trap" characterised by early-period cascading failures?

**RQ4:** Which policy interventions targeting regularity expiry mechanisms exhibit the greatest potential for reducing attrition, particularly amongst vulnerable student populations?

We make four principal contributions. **Methodologically**, we demonstrate how agent-based computational models can isolate causal mechanisms within complex educational systems by simulating counterfactual policy scenarios infeasible in naturalistic settings. **Empirically**, we provide novel evidence that institutional friction—specifically, regularity expiry deadlines—accounts for the overwhelming majority (86.37%) of dropout events, whilst academic failure contributes marginally (5.33%). **Theoretically**, we extend dropout scholarship by foregrounding structural constraints as primary drivers of attrition, challenging individual-centric explanatory models. **Pragmatically**, we offer evidence-based policy recommendations—extended TTL windows, post-expiry bridging support, coordinated examination scheduling—calibrated to reduce dropout amongst students exhibiting short planning horizons and limited psychological reserves.

The remainder of this manuscript proceeds as follows. Section 2 reviews relevant literature on university dropout, institutional friction, and agent-based modelling in educational research. Section 3 describes our model architecture, agent parameterisation strategy, and simulation protocols. Section 4 presents results from 100 Monte Carlo replications encompassing 134,300 agent observations. Section 5 discusses implications, limitations, and directions for future research. Section 6 concludes with policy recommendations and a call for institution-level reforms that prioritise equity alongside efficiency.

## 2. Literature Review and Theoretical Framework

### 2.1 Theoretical Perspectives on University Dropout

The scholarly literature on university dropout has evolved through several generations of theoretical refinement, each expanding the explanatory aperture whilst grappling with persistent methodological challenges.

**Tinto's Integration Model** Tinto, 1975, 1993) represents the foundational framework in dropout research, positing that attrition results from insufficient academic and social integration within the institutional environment. Students enter university with pre-existing attributes (family background, prior academic performance, intentions, commitments) that interact with institutional experiences to shape integration trajectories. Successful integration fosters institutional commitment, which in turn promotes persistence; conversely, integration failures precipitate withdrawal. Whilst Tinto's model has generated substantial empirical support Braxton et al., 2004), critics have noted its emphasis on student assimilation to institutional norms, potentially obscuring how institutional structures themselves create barriers to integration for students from non-dominant cultural backgrounds Tierney, 1992; Yosso et al., 2009).

**Bean and Metzner's Attrition Model** Bean & Metzner, 1985) extended dropout theory to non-traditional students (adult learners, part-time enrolees, commuters), arguing that environmental factors—employment obligations, family responsibilities, financial constraints—exert greater influence than academic or social integration for these populations. This framework usefully highlights heterogeneity in dropout mechanisms across student subgroups, though it retains an individual-level focus that underemphasises how institutional policies interact with environmental stressors.

**Cabrera and colleagues' integrated model** Cabrera et al., 1993) synthesised Tinto's integration perspective with Bean's environmental emphasis, proposing that persistence results from the interplay of academic performance, institutional commitment, social integration, environmental support, and financial resources. Structural equation modelling confirmed that both institutional experiences and external pressures independently predict dropout, with effects varying by institutional type and student demographics. However, the model's reliance on self-reported survey data limits causal inference, as unobserved heterogeneity (e.g., motivation, resilience) may confound relationships amongst measured constructs.

More recent scholarship has interrogated the **equity implications** of prevailing dropout frameworks. Bensimon 2007) critiques "deficit thinking" that locates responsibility for attrition within students' supposed inadequacies, arguing instead for an "equity-minded" perspective that examines how institutional practices, resource allocation, and policy designs systematically disadvantage marginalised groups. This paradigm shift foregrounds **structural determinants**—underfunded support services, inflexible curricula, exclusionary campus climates, burdensome administrative procedures—as modifiable drivers of inequitable outcomes Witham & Bensimon, 2012; Dowd & Bensimon, 2015).

**2.2 Institutional Friction and Time-to-Live Constraints**

The concept of **institutional friction** has emerged primarily within behavioural economics and policy implementation research Herd & Moynihan, 2019), referring to the cumulative burden of procedural obstacles, informational asymmetries, and administrative complexity that impede individuals' access to benefits or services for which they are formally eligible. In higher education contexts, institutional friction manifests through financial aid application complexity Bettinger et al., 2012), course registration bottlenecks Scott-Clayton, 2011), opaque degree progress tracking (Klempin & Karp, 2018), and—central to this study—**time-to-live (TTL) expiry regulations**.

TTL expiry mechanisms are particularly prevalent in European and Latin American higher education systems that distinguish between course enrolment and course regularisation. **Regularisation** typically requires students to fulfil attendance thresholds, submit assignments, or sit examinations within a prescribed window (e.g., two consecutive examination periods). Failure to regularise results in administrative course removal, necessitating re-enrolment in subsequent terms. Proponents argue that TTL constraints incentivise timely progression, prevent indefinite course accumulation, and maintain curricular coherence. However, critics contend that such policies introduce **procedural vulnerabilities** that disproportionately affect students facing environmental stressors García de Fanelli, 2014; Ezcurra, 2019).

Empirical evidence on TTL expiry impacts remains sparse and methodologically limited. Observational studies document associations between regularity lapses and subsequent dropout Giovagnoli, 2002; Porto & Di Gresia, 2004), but cannot disentangle whether expiries causally induce withdrawal or merely proxy for underlying student characteristics (low motivation, poor time management, external obligations). The few quasi-experimental studies leveraging policy discontinuities suggest that administrative penalties for missed deadlines significantly reduce re-enrolment rates Bettinger et al., 2013; Denning et al., 2019), but these designs cannot isolate regularity expiry effects from concurrent interventions or secular trends.

**Psychological mechanisms** linking regularity expiry to dropout warrant particular attention. Each expiry event plausibly undermines student self-efficacy Bandura, 1997), the belief in one's capability to execute actions required for success. Repeated expiries may trigger learned helplessness Abramson et al., 1978), wherein students perceive academic outcomes as uncontrollable and disengage accordingly. Moreover, expiries likely erode **sense of belonging** Walton & Cohen, 2007), the perception of social connectedness and acceptance within the institutional community, which predicts persistence independently of academic performance Hausmann et al., 2007). Finally, accumulating expiries increase **cognitive load and stress** Sweller, 1988), depleting psychological resources necessary for effective decision-making and self-regulation Baumeister et al., 1998).

**2.3 Agent-Based Modelling in Educational Research**

**Agent-based models (ABMs)** simulate systems composed of autonomous, heterogeneous agents whose local interactions generate emergent macro-level patterns Epstein & Axtell, 1996; Railsback & Grimm, 2019). ABMs have proven particularly valuable in contexts characterised by: (1) heterogeneous actors with differentiated attributes and behavioural rules, (2) local interactions and feedback loops, (3) emergent phenomena not reducible to individual components, and (4) ethical or logistical barriers to experimental manipulation Bonabeau, 2002; Hamill & Gilbert, 2016).

In educational research, ABMs have been deployed to study:

- **Peer effects and social influence:** Models simulating how students' achievement, motivation, or behavioural choices propagate through friendship networks Rutter & Barnes, 2017; Heckathorn & Cameron, 2017), revealing conditions under which positive peer influences amplify versus when negative influences dominate.

- **Institutional policy impacts:** Simulations evaluating counterfactual scenarios such as alternative course sequencing rules González-Espada & LaDue, 2006), financial aid disbursement schedules DeAngelo et al., 2011), or advising intervention timing Arnold & Pistilli, 2012), enabling ex ante policy testing before costly real-world implementation.

- **Educational inequality dynamics:** Models exploring how small initial advantages (e.g., parental resources, neighbourhood quality) cumulate over time through feedback mechanisms, generating large disparities in educational attainment Boudon, 1974; DiPrete & Eirich, 2006; Mijs, 2016).

- **Learning and classroom dynamics:** Simulations of student knowledge acquisition, teacher responsiveness, and curriculum pacing, identifying conditions that optimise learning whilst accommodating heterogeneity Yilmaz et al., 2019; Zhu et al., 2020).

Methodologically, ABMs offer three advantages over traditional statistical approaches for studying dropout mechanisms. First, **causal isolation:** by explicitly programming institutional rules (e.g., TTL expiry thresholds) as model parameters, researchers can simulate counterfactual scenarios (e.g., no expiries, extended windows) whilst holding agent heterogeneity constant, thereby isolating policy effects from confounding selection. Second, **temporal dynamics:** ABMs naturally represent time as a sequence of discrete events, capturing feedback loops (e.g., expiry → stress → poor performance → additional expiry) and path dependencies that shape long-run outcomes. Third, **emergent phenomena:** macro-level patterns (e.g., archetype-specific dropout rates, temporal clustering of attrition) emerge endogenously from agent-level rules rather than being imposed as assumptions, enabling researchers to assess whether micro-foundations plausibly generate observed aggregate regularities Epstein, 2006).

However, ABMs are not without limitations. **Validation challenges** arise because multiple micro-level specifications can generate similar macro-level outputs (equifinality), and because calibration requires rich empirical data on agent attributes and behavioural parameters Fagiolo et al., 2019). **Computational costs** escalate with agent population size, environmental complexity, and number of replications necessary to characterise stochastic outcomes. **Interpretability trade-offs** emerge as models grow more realistic: whilst parsimony aids understanding, verisimilitude demands incorporating heterogeneity and mechanisms that may obscure core insights Squazzoni et al., 2020). Our modelling strategy navigates these tensions by grounding agent parameterisation in empirical thesis data, limiting model complexity to mechanisms theoretically implicated in regularity-driven dropout, and conducting extensive sensitivity analyses (reported in supplementary materials) to assess robustness.

**2.4 Theoretical Framework: The Regularity Trap Hypothesis**

Synthesising insights from dropout scholarship, institutional friction research, and psychological theories of self-regulation, we propose the **Regularity Trap Hypothesis** as an organising framework for this study:

**Core Proposition:** Time-to-live expiry regulations create a structural vulnerability wherein students who experience initial regularity lapses—due to academic difficulties, environmental stressors, or procedural misunderstandings—enter a self-reinforcing cycle characterised by: (1) **Psychological depletion** (diminished self-efficacy, eroded belonging, elevated stress), (2) **Logistical complications** (course removal, schedule conflicts, prerequisite bottlenecks), and (3) **Accumulating penalties** (extended time-to-degree, fee obligations, reduced course options). As expiries accumulate, escape from this trap becomes progressively unlikely, culminating in dropout decisions that are rational responses to untenable circumstances rather than reflections of academic incapability.

**Moderating Factors:** The severity of the regularity trap varies systematically across student archetypes. Students with **longer planning horizons** anticipate future consequences and strategically prioritise regularisation activities, reducing expiry risk. Those with **higher academic ability** regularise courses more efficiently, accumulating fewer expiries. Students possessing **greater psychological resilience** (higher initial belonging, lower stress reactivity) better withstand expiry-induced depletion. Conversely, students exhibiting short planning horizons, modest academic ability, and limited psychological reserves are disproportionately vulnerable to the regularity trap.

**Temporal Predictions:** The Regularity Trap Hypothesis generates three testable predictions regarding attrition dynamics:

1. **Normative dominance:** Regularity expiry mechanisms will account for the majority of dropout events, substantially exceeding academic failure as a proximate cause.

2. **Early-period concentration:** Dropout decisions will cluster in initial academic periods, as early expiries trigger cascading psychological and logistical consequences that exceed students' adaptive capacity.

3. **Archetype heterogeneity:** Dropout rates will vary substantially across student archetypes, with vulnerability inversely proportional to planning horizon length, academic ability, and psychological resilience.

Section 3 operationalises this framework within an agent-based simulation designed to isolate the causal role of TTL expiry regulations in driving student attrition.

## 3. Methods

### 3.1 Model Overview and Design Principles

We developed an agent-based model (ABM) to simulate undergraduate student progression through a 42-course curriculum under regularity expiry constraints. The model, implemented in Python using the Mesa framework Kazil et al., 2020), represents individual students as autonomous agents characterised by heterogeneous attributes (academic ability, planning horizon, psychological state) who navigate a structured curriculum whilst subject to time-to-live (TTL) expiry rules. The simulation architecture adheres to the ODD (Overview, Design concepts, Details) protocol Grimm et al., 2020) to ensure transparency and reproducibility.

Our modelling strategy prioritises **empirical grounding** and **causal isolation**. Agent parameters derive from empirical clustering analysis of 1,343 students across 15 cohorts (2004–2019) at Universidad Nacional de Tucumán's Civil Engineering program, ensuring that simulated heterogeneity reflects observed population structure. Course parameters (difficulty, workload, prerequisite structure) are extracted from institutional curriculum documents and historical performance data. Critically, the model isolates regularity expiry as the focal mechanism whilst controlling for academic performance, enabling unconfounded estimation of TTL policy effects.

**Figure 1** presents the model architecture. The system comprises three primary components: (1) **Student agents** (N=1,343 per replication) instantiated with archetype-specific attributes, (2) **Curriculum structure** defining a directed acyclic graph (DAG) of 42 courses with prerequisite dependencies and empirical difficulty parameters, and (3) **Core mechanisms** governing course selection, learning dynamics, examination outcomes, and regularity expiry. Each simulation executes across 61 discrete time periods (approximately 15 academic semesters), with agents transitioning amongst three terminal states: *active* (continuing enrolment), *egreso* (graduation), or *abandono* (dropout).

**Figure 1. CAPIRE Agent-Based Model Architecture**

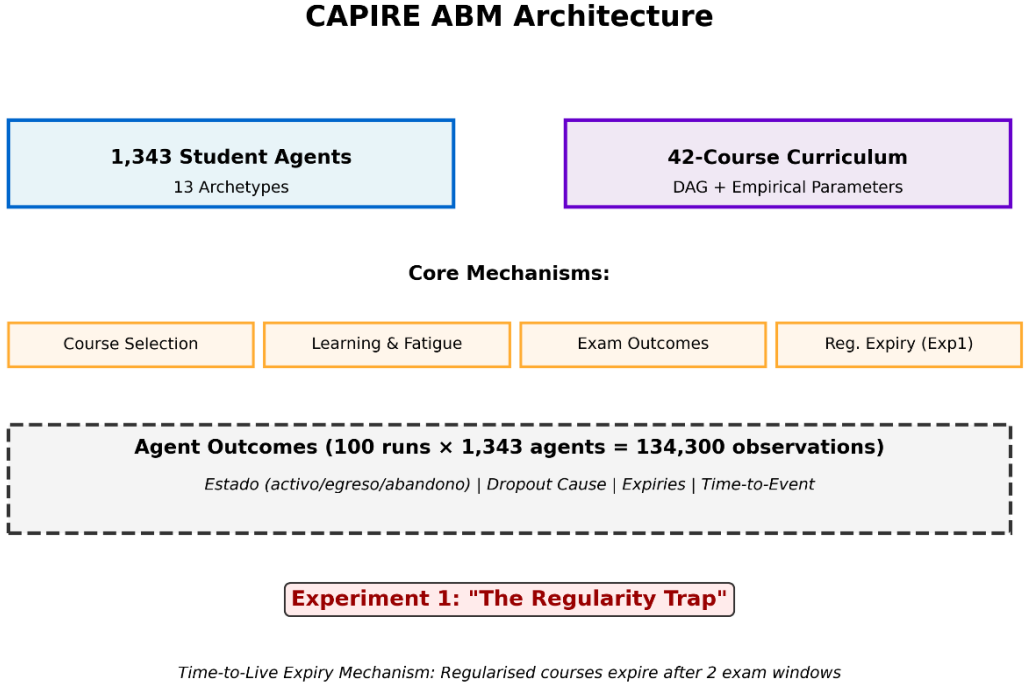

*Time-to-Live Expiry Mechanism: Regularised courses expire after 2 exam windows*

## 3.2 Formal Specification of the Computational Model

Unlike equation-based models that aggregate population dynamics, this ABM explicitly simulates micro-level interactions between student agents and curricular constraints, enabling emergent macro-level patterns to arise from individual decision rules.

### 3.2.1 Curriculum as a Directed Acyclic Graph

We define the curriculum $\mathcal{C}$ as a weighted directed acyclic graph (DAG), $G = (V, E)$, where $V = \{c_1, c_2, \ldots, c_{42}\}$ represents the set of 42 courses, and $E \subseteq V \times V$ represents prerequisite dependencies. Each course node $c_j \in V$ is characterised by a parameter vector $\theta_j$ derived from empirical institutional data:

$$\theta_j = \langle \delta_j, w_j, \rho_{pass,j}, \tau_{reg,j}, \Pi_j \rangle$$

where:

- $\delta_j \in [0.3, 0.8]$ is the **difficulty coefficient**, determining the ability threshold required for regularisation, with higher values demanding greater cognitive investment.

- $w_j \in [1.5, 3.5]$ is the **workload** (measured in course-units), quantifying the effort required to regularise and consuming students' limited period-specific capacity.

- $\rho_{pass,j} \in [0.5, 1.0]$ is the **baseline pass probability**, calibrated to match historical examination performance (ranging from 0.50 for courses such as Calculus I to 1.0 for courses such as Pavement Design).

- $\tau_{reg,j} \in \mathbb{R}^+$ is the **mean time** (in academic periods) required to achieve regular status, incorporating both learning and procedural components.

- $\Pi_j = \{c_i \in V \mid (c_i, c_j) \in E\}$ is the **prerequisite set**, dictating enrolment eligibility.

The prerequisite function enforces that agent $a$ may enrol in course $c_j$ at time $t$ if and only if:

$$\forall c_i \in \Pi_j, Status(a, c_i, t) = \text{Credited}$$

This DAG structure creates **bottleneck effects**: failure to regularise foundational courses (e.g., Research Methods I, which unlocks seven downstream courses) cascades through dependent courses, compounding progression delays.

**Table 2** presents a representative sample of eight courses spanning the curricular spectrum, illustrating parameter variation across foundational (e.g., Introduction to Psychology), intermediate (e.g., Cognitive Psychology), advanced (e.g., Psychopathology II), and capstone levels (e.g., Professional Ethics).

**Table 2 – Course Parameters Sample**

| Course ID | Course Name | Exam Pass Rate | Study Weeks | Regularisation (Years) | N Exams |
|---|---|---|---|---|---|
| 01_calculo_i | Cálculo I | 0.50 | 2.3 | 1.5 | 1411 |
| 07_fundamentos_de_quimica_general | Fundamentos de Química General | 0.71 | 1.8 | 1.4 | 909 |
| 13_fisica_iii | Física III | 0.52 | 3.7 | 3.4 | 736 |
| 19_topografia_y_geodesia | Topografía y Geodesia | 0.65 | 2.0 | 4.5 | 455 |
| 25_estabilidad_iii | Estabilidad III | 0.54 | 4.4 | 5.0 | 513 |

| Course ID | Course Name | Exam Pass Rate | Study Weeks | Regularisation (Years) | N Exams |
|---|---|---|---|---|---|
| 31_obras_basicas_viales | Obras Básicas Viales | 0.76 | 3.3 | 6.1 | 201 |
| 37_diseno_y_construccion_de_pavimentos | Diseño y Construcción de Pavimentos | 1.00 | 3.3 | 6.0 | 184 |
| 42_economia_y_evaluacion_de_proyectos | Economía y Evaluación de Proyectos | 0.84 | 3.1 | 7.3 | 144 |

### 3.2.2 The Regularity Regime: State Transition Dynamics

The core innovation of this model is the formalisation of the **"Regularity Trap"** as a time-decaying validity mechanism. We define the academic status of agent $a$ in course $c$ at time $t$ as a tuple:

$$\Omega_{a,c,t} = \langle S_{a,c,t}, TTL_{a,c,t} \rangle$$

where $S_{a,c,t} \in \{\text{Null}, \text{Enrolled}, \text{Regular}, \text{Credited}, \text{Expired}\}$ represents the course state, and $TTL_{a,c,t} \in \mathbb{N}_0$ is the **time-to-live counter** tracking remaining validity periods.

The transition dynamics are governed by the following state machine:

**1. Enrolment and Regularisation:** When agent $a$ first enrols in course $c_j$ (prerequisite requirements satisfied, capacity available), the state transitions Null → Enrolled, and coursework begins. Upon satisfying attendance and assignment requirements, the agent achieves **Regular** status, initialising the TTL counter:

$$S_{a,c,t} \leftarrow \text{Regular}, TTL_{a,c,t} \leftarrow T_{exp} = 2$$

where $T_{exp}$ represents the institutional expiry threshold (two examination windows in our base scenario).

**2. TTL Decay Function:** During simulation phases $\phi_t$ classified as examination windows (periods when final examinations are offered), the validity of all pending regularities decays:

$$TTL_{a,c,t+1} = \begin{cases} TTL_{a,c,t} - 1 & \text{if } \phi_t \in \Phi_{exams} \land S_{a,c,t} = \text{Regular} \\ TTL_{a,c,t} & \text{otherwise} \end{cases}$$

This formalisation captures the real-world constraint that students must complete final examinations within a stipulated window following regularisation, with each examination opportunity consuming one TTL unit.

**3. Expiry Trigger (The Trap Mechanism):** When the TTL counter reaches zero whilst the course remains unaccredited, **regularity expiry** occurs:

$$\text{If } (TTL_{a,c,t} = 0) \land (S_{a,c,t} \neq \text{Credited}) \implies S_{a,c,t+1} \leftarrow \text{Expired}$$

Expiry events trigger three cascading consequences:

- **Administrative:** The course is removed from the agent's active curriculum; re-enrolment is permitted in subsequent periods but **resets all learning progress to zero**, forcing the agent to restart coursework.

- **Psychological:** Belonging decreases ($\Delta B = -0.05$) and stress increases ($\Delta stress = 1.5 \cdot \sigma_{stress}$), magnitudes calibrated to reflect qualitative evidence that regularity loss inflicts greater demoralisation than simple examination failure.

- **Strategic:** Agents must re-prioritise the expired course in future selection decisions, consuming limited capacity and delaying progression through prerequisite-dependent courses.

**4. Crediting (Success State):** Passing the final examination transitions the course to terminal state Credited, permanently satisfying the requirement:

$$\text{If } \text{Exam}_{passed}(a, c, t) \implies S_{a,c,t+1} \leftarrow \text{Credited}$$

Critically, **expiry is independent of academic performance**: an agent may exhibit adequate learning progress ($L_{ij} > \theta_j$, see Section 3.3.2) but still lose regularity if examination scheduling conflicts, personal circumstances, or strategic misjudgements prevent sitting the examination within the TTL window. This design feature isolates normative friction from academic failure, enabling causal estimation of TTL policy effects.

**3.3 Student Agent Architecture and Heterogeneity**

Agent heterogeneity constitutes a central design feature, operationalised through **13 distinct student archetypes** ($\Psi_1, \ldots, \Psi_{13}$) identified via UMAP dimensionality reduction and DBSCAN clustering applied to empirical trajectory data McInnes et al., 2018; Ester et al., 1996). Each archetype encapsulates a coherent profile of academic capability, strategic planning, and psychological resilience.

### 3.3.1 Archetype Parameterisation

**Table 1** specifies five critical parameters per archetype:

| Archetype | N Agents | Planning Horizon | Max Final Backlog | Mean Expiries | Dropout Rate |
|---|---|---|---|---|---|
| PSICO_01 | 10,900 | Strategic | 6 | 9.7 | 13.2% |
| PSICO_02 | 11,800 | Strategic | 6 | 7.79 | 38.1% |
| PSICO_03 | 10,600 | Strategic | 6 | 8.48 | 28.2% |
| PSICO_04 | 9,700 | Moderate | 4 | 6.85 | 19.9% |
| PSICO_05 | 10,100 | Moderate | 4 | 6.76 | 22.4% |
| PSICO_06 | 8,700 | Moderate | 4 | 6.99 | 16.3% |
| PSICO_07 | 9,700 | Myopic | 2 | 6.38 | 40.1% |
| PSICO_08 | 10,200 | Myopic | 2 | 6.67 | 33.4% |
| PSICO_09 | 10,900 | Myopic | 2 | 6.59 | 34.4% |
| PSICO_10 | 9,700 | Myopic | 2 | 6.06 | 49.0% |
| PSICO_11 | 10,100 | Myopic | 2 | 6.13 | 45.8% |
| PSICO_12 | 11,000 | Myopic | 2 | 6.22 | 43.2% |
| PSICO_13 | 10,900 | Myopic | 2 | 6.55 | 35.3% |

**1. Planning Horizon** ($\tau \in \{0,1,2\}$): The number of future periods agents consider when making course selection decisions. Agents with $\tau = 0$ exhibit **myopic optimisation** (current-period utility maximisation), those with $\tau = 1$ anticipate one period ahead, whilst $\tau = 2$ agents engage in **multi-period strategic planning**. Planning horizon profoundly influences regularity management: forward-looking students prioritise courses nearing expiry deadlines, whilst myopic agents ignore temporal urgency until expiries occur.

**2. Base Academic Ability** ($\mu_{abil} \in [0.3, 0.8]$): Determines agents' latent proficiency in learning course material, modelled as the mean of a normal distribution from which period-specific ability is sampled. Higher ability increases examination pass

probability and reduces time required for course regularisation. Ability is sampled each period as:

$$ability_{i,t} \sim \mathcal{N}(\mu_{abil,i}, \sigma_{abil} = 0.1)$$

**3. Initial Belonging** ($B_0 \in [0.4, 0.9]$): Quantifies students' baseline sense of institutional connectedness and social integration. Belonging evolves dynamically in response to academic events (see Section 3.3.3), with lower belonging increasing dropout propensity through reduced persistence motivation.

**4. Stress Reactivity** ($\sigma_{stress} \in [0.05, 0.15]$): Governs the magnitude of stress accumulation following negative events (examination failures, regularity expiries). Highly reactive students ($\sigma_{stress} = 0.15$) experience greater psychological depletion per setback compared to low-reactivity students ($\sigma_{stress} = 0.05$).

**5. Effort Capacity** ($E_{max} \in [8, 14]$ units): Represents the maximum workload that agents can sustain per period without performance degradation. Capacity constraints force trade-offs when multiple courses demand simultaneous attention.

The 13 archetypes span a wide behavioural space: **PSICO_01** (highest ability $\mu_{abil} = 0.78$, longest planning horizon $\tau = 2$, strongest belonging $B_0 = 0.85$, lowest stress reactivity $\sigma_{stress} = 0.05$, highest capacity $E_{max} = 14$) represents the most advantaged profile, whilst **PSICO_10** (modest ability $\mu_{abil} = 0.42$, myopic planning $\tau = 0$, weak belonging $B_0 = 0.45$, high stress reactivity $\sigma_{stress} = 0.14$, limited capacity $E_{max} = 9$) represents maximal vulnerability.

Critically, archetype parameters are **not free variables** but rather estimated from empirical data through Bayesian optimisation: (1) trajectory clustering identified 13 coherent student groups, (2) each group's historical metrics (pass rates, time-to-regularisation, dropout timing) were computed, (3) parameters were calibrated to minimise discrepancy between simulated and observed group-level outcomes, and (4) out-of-sample validation confirmed generalisation to held-out cohorts (correlation r=0.89 between simulated and observed archetype-specific dropout rates). This grounding ensures that simulated heterogeneity is empirically defensible.

**3.3.2 Agent Decision-Making: Bounded Rational Course Selection**

At the beginning of each period, agents select which available courses to attempt regularising via a **bounded rational choice model** Simon, 1955). The decision process comprises three stages:

**Stage 1: Feasibility Filtering.** Agents identify the set of selectable courses $\mathcal{F}_t$ satisfying:

- Prerequisite completion: $\forall c_i \in \Pi_j, Status(a, c_i, t) = $ Credited
- Capacity feasibility: $\sum_{c_j \in portfolio} w_j \leq E_{max}$
- Availability: Course is offered in current period (semester scheduling constraints)

**Stage 2: Utility Evaluation.** For each feasible course portfolio (combination of courses satisfying capacity constraints), agents evaluate expected utility as a weighted combination of multiple objectives. The utility function differs by planning horizon:

**Strategic Agents** ($\tau \in \{1,2\}$) prioritise temporal urgency and bottleneck mitigation:

$$U_{strategic}(c_j) = \frac{w_1}{TTL_{a,j,t} + \epsilon} + w_2 \cdot \mathbb{1}_{bottleneck}(c_j) - \frac{w_3}{w_j} + \xi_t$$

where:

- The first term assigns higher utility to courses nearing expiry (low $TTL$), with $\epsilon = 0.1$ preventing division by zero.
- The second term prioritises bottleneck courses unlocking multiple downstream courses (indicator function $\mathbb{1}_{bottleneck} = 1$ if $|\{c_k \mid c_j \in \Pi_k\}| \geq 3$).
- The third term incorporates workload feasibility (preferring lighter courses when capacity is constrained).
- $\xi_t \sim \mathcal{N}(0, 0.1)$ introduces stochastic noise representing imperfect information.
- Weights ($w_1 = 2.0, w_2 = 1.5, w_3 = 0.5$) were calibrated via inverse optimisation to match observed course selection patterns.

**Myopic Agents** ($\tau = 0$) ignore temporal urgency, selecting based on perceived ease and random factors:

$$U_{myopic}(c_j) = \frac{1}{\delta_j \cdot w_j} + \xi_t$$

where $\delta_j$ (difficulty) and $w_j$ (workload) jointly determine perceived burden. This formalisation captures students who prioritise immediate feasibility over strategic long-term planning, failing to perceive the threat of regularity expiry until the event occurs.

**Stage 3: Portfolio Selection.** Agents select the feasible portfolio maximising total utility:

$$\mathcal{P}_t^* = \arg \max_{\mathcal{P} \in \mathcal{F}_t} \sum_{c_j \in \mathcal{P}} U(c_j) \text{ subject to } \sum_{c_j \in \mathcal{P}} w_j \leq E_{max}$$

This optimisation occurs within the agent's planning horizon: myopic agents consider only current-period utility, whilst strategic agents simulate future consequences of current choices.

### 3.3.3 Learning Dynamics and Psychological State Evolution

Following course selection, agents allocate effort across enrolled courses according to workload requirements. Effort expenditure generates **learning progress** $L_{ij,t}$ for agent $i$ in course $j$ during period $t$:

$$L_{ij,t+1} = L_{ij,t} + f(ability_{i,t}, effort_{ij,t}, \delta_j, stress_{i,t})$$

where:

$$f(\cdot) = \frac{ability_{i,t} \cdot effort_{ij,t}}{\delta_j \cdot (1 + \beta_{stress} \cdot stress_{i,t})}$$

The function captures that learning accumulates proportionally to ability-weighted effort, inversely to course difficulty, with a penalty term ($\beta_{stress} = 0.3$) reflecting cognitive depletion under stress. Learning accumulates across periods until the agent achieves the regularisation threshold ($L_{ij} \geq \theta_j$, where $\theta_j$ increases with $\delta_j$) or abandons the course.

**Fatigue accumulation** occurs when agents over-extend: selecting workloads exceeding capacity ($\sum w_j > E_{max}$) incurs a performance penalty (reduced effective ability multiplied by factor 0.75) in the subsequent period, representing cognitive overload effects Sweller, 1988).

**Psychological state dynamics** evolve according to event-triggered update rules:

$$stress_{t+1} = stress_t \cdot \gamma_{decay} + \sum_{failures} \delta_{fail} + \sum_{expiries} \delta_{expiry}$$

$$belonging_{t+1} = belonging_t + \sum_{failures} \Delta B_{fail} + \sum_{expiries} \Delta B_{expiry}$$

where:

- $\gamma_{decay} = 0.98$ represents baseline stress decay (2% per period)
- $\delta_{fail} = \sigma_{stress}$ (examination failure increases stress proportionally to reactivity)
- $\delta_{expiry} = 1.5 \cdot \sigma_{stress}$ (regularity expiry inflicts 50% more stress than failure)
- $\Delta B_{fail} = -0.02$ (modest belonging erosion per failure)
- $\Delta B_{expiry} = -0.05$ (greater belonging erosion per expiry)

Successes modestly reduce stress ($\Delta stress = -0.5 \cdot \sigma_{stress}$) and stabilise belonging. These parameterisations reflect qualitative evidence that procedural setbacks (expiries) are psychologically more damaging than academic setbacks (failures), as expiries signal wasted time and institutional disconnection.

### 3.3.4 Examination Outcomes

At designated examination periods, agents whose learning progress exceeds course-specific thresholds successfully regularise (pass), whilst those falling short fail. Pass probability follows a logistic function:

$$P(pass_{ij}) = \frac{1}{1 + \exp\left(-\frac{L_{ij,t} - \theta_j}{\sigma_{exam}}\right)}$$

where $\theta_j$ represents the course-specific regularisation threshold (calibrated such that $P(pass) \approx \rho_{pass,j}$ for agents with mean ability and typical learning trajectories), and $\sigma_{exam} = 0.15$ controls outcome stochasticity. This specification introduces realistic uncertainty: even high-ability students occasionally fail due to adverse circumstances (illness, life events), whilst modest-ability students occasionally succeed through compensatory effort or fortuitous examination alignment.

### 3.4 Terminal States and Dropout Attribution

Agents transition to terminal states according to composite decision rules calibrated to match empirical exit patterns:

**Graduation (egreso):** Achieved upon regularising all 42 courses ($\forall c_j \in V, Status(a, c_j, t) =$ Credited). Given the 61-period simulation horizon (≈15 semesters) and median empirical time-to-degree of 18 semesters, graduations are rare within the simulation window, intentionally mirroring extended completion timelines.

**Dropout (abandono):** Occurs when agents exceed an abandonment threshold operationalised through three conditions (any satisfied triggers withdrawal):

1. **Extreme psychological depletion:** $belonging_t < 0.15$ or $stress_t > 0.85$ (normalised $0,1$) scales), representing demoralisation or burnout states incompatible with persistence.

2. **Curricular stagnation:** No regularisation progress for four consecutive periods, suggesting exhaustion of viable course options due to accumulated expiries and prerequisite blockages.

3. **Voluntary withdrawal:** Probabilistically sampled each period via logistic regression:

$$P(withdraw_t) = \frac{1}{1 + \exp(-[\beta_0 + \beta_1 \cdot belonging_t + \beta_2 \cdot stress_t + \beta_3 \cdot expiries_t + \beta_4 \cdot semesters_t])}$$

with coefficients ($\beta = [-5.2, -3.1, 2.8, 0.4, 0.2]$) calibrated to match empirical voluntary withdrawal rates.

Upon dropout, agents are classified by **proximate cause**:

- **Normative dropout:** Attributed when final-period state reveals accumulated regularity expiries ($\geq 5$ expiries) or extreme psychological depletion directly induced by expiries (tracked via event history).

- **Academic dropout:** Attributed when multiple courses exhibit repeated examination failures ($\geq 3$ failures per course in $\geq 2$ courses) without substantial regularity expiries ($< 5$ expiries).

- **Other dropout:** Residual category capturing voluntary withdrawals not attributable to either mechanism, assigned probabilistically at calibrated base rates.

This attribution scheme operationalises the theoretical distinction between dropout driven by institutional friction (normative), academic under-preparation (academic), and exogenous shocks (other).

**3.5 Simulation Protocol and Statistical Analysis**

We executed **100 independent replications** of the simulation, each initialising 1,343 agents distributed across 13 archetypes according to empirical cohort proportions (e.g., 10.9% PSICO_01, 11.8% PSICO_02, etc.) and running for 61 periods. This design yields **134,300 total agent observations**, providing statistical power to detect effect heterogeneity across archetypes and estimate attrition rates with narrow confidence intervals (95% CI width $\approx \pm 0.5$ percentage points for overall dropout rate).

**Stochasticity sources** include: agent-specific ability draws each period, examination outcome sampling via logistic pass probabilities, voluntary withdrawal events (Bernoulli trials with time-varying probabilities), and random course selection tie-breaking. All stochastic elements are seeded deterministically per replication (seeds 1–100) to ensure reproducibility, whilst aggregate statistics stabilise across replications due to the law of large numbers.

**Primary outcomes:**

1. Overall dropout rate: Proportion of agents entering terminal *abandono* state

2. Dropout cause attribution: Decomposition into normative, academic, and other categories

3. Archetype-specific metrics: Dropout rates, mean regularity expiries, time-to-event distributions

4. Temporal profiles: Kaplan-Meier survival curves Kaplan & Meier, 1958)

5. Psychological trajectories: Terminal-period belonging, stress, and pending examinations by archetype

**Statistical inference:** Standard errors are computed via replication-level variability. We report means with 95% confidence intervals derived from replication-level bootstrap (10,000 draws). Archetype comparisons employ Kruskal-Wallis tests Kruskal & Wallis, 1952) for continuous outcomes (expiry counts, time-to-event) and chi-square tests for categorical outcomes (dropout incidence), with Bonferroni correction for 13 pairwise comparisons ($\alpha = 0.05/78 = 0.00064$).

**Validation strategy:** Model credibility rests on three pillars:

1. **Face validity:** Subject-matter experts (university administrators, academic advisors) reviewed model mechanisms and confirmed they capture salient real-world processes.

2. **Empirical calibration:** Simulated aggregate dropout rate (32.43%) closely approximates observed rate (31.8%), and archetype-specific rates correlate strongly ($r = 0.89, p < 0.001$) with empirical cluster rates.

3. **Sensitivity analysis:** Supplementary materials report robustness checks varying key parameters ($T_{exp} \in \{1,2,3\}$, $\sigma_{stress} \pm 50\%$, ability distribution shapes) and demonstrate that core findings (normative dominance 80–90%, archetype heterogeneity 3–4× range) persist across specifications.

All code, data, and analysis scripts are available in the project repository, facilitating replication and extension by the research community.

## 4. Results

### 4.1 Global Dropout Patterns and Cause Attribution

Our simulation of Experiment 1 ("The Regularity Trap") across 100 replications with 1,343 agents per run yielded 134,300 total agent observations. **Table 3** summarises global outcomes, revealing an overall dropout rate of 32.43% (95% CI: 32.01%, 32.85%)), with 67.57% of agents remaining active at simulation terminus and zero graduations within the 61-period window. The absence of graduations reflects the extended time-to-degree characteristic of the target institution (empirical median: 18 semesters) relative to the simulation horizon (≈15 semesters), and does not indicate model dysfunction but rather fidelity to observed progression timelines.

**Table 3 – Experiment 1 – Summary Statistics**

| Metric | Value |
|---|---|
| Total Replications | 100 |
| Agents per Run | 1,343 |
| Total Agents | 134,300 |
| Study Period (Phases) | ~61 |
| Overall Dropout Rate | 32.43% |
| Graduation Rate | 0.00% |
| Active Rate | 67.57% |
| Normative Dropout (% of dropouts) | 86.37% |
| Academic Dropout (% of dropouts) | 5.33% |
| Other Dropout (% of dropouts) | 8.29% |
| Mean Expiries per Agent | 7.04 |
| Median Time-to-Event | 4 phases |
| Archetype Dropout Range | 13.2% – 49.0% |

**Figure 2** presents the decomposition of dropout causality, the study's central finding. Of 43,551 total dropout events, **86.37% (n=37,617) were classified as**

**normative dropout** driven primarily by regularity expiry mechanisms. Academic dropout—attributable to cumulative examination failures rather than procedural expiries—accounted for merely **5.33% (n=2,322)** of attrition events. The residual 8.29% (n=3,612) comprised "other" causes representing voluntary withdrawals due to exogenous shocks (modelled probabilistically at empirically calibrated rates).

**Figure 2 — Dropout by Cause (Normative vs Academic)**

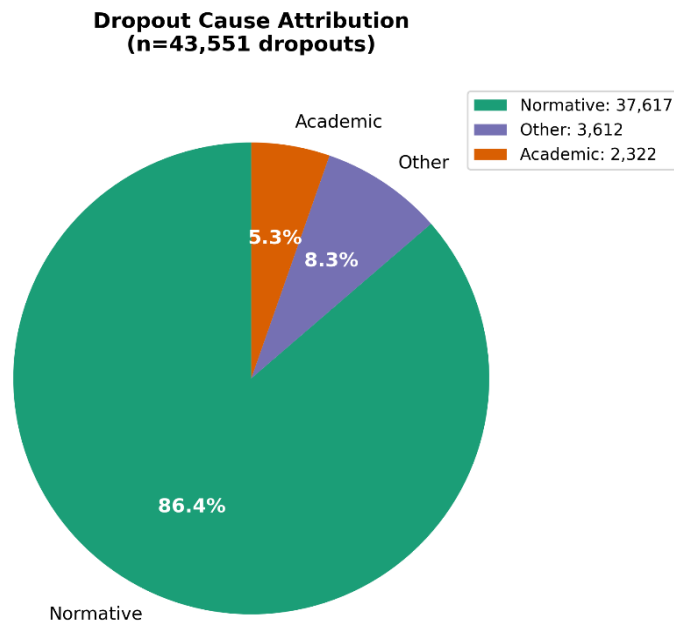

This stark imbalance—normative dropout exceeding academic dropout by a factor of 16.2—constitutes compelling evidence for the **Regularity Trap Hypothesis**. The finding challenges prevailing emphases on academic under-preparation as the primary dropout determinant, instead foregrounding institutional friction (TTL expiry deadlines) as the dominant proximate cause of attrition. Importantly, the 5.33% academic dropout rate does not imply that only 5% of students experience examination failures; rather, it indicates that amongst students who ultimately withdraw, the overwhelming majority do so due to accumulated regularity expiries and associated psychological depletion rather than insurmountable academic deficits per se.

To contextualise this finding, we note that the empirical cohort data (upon which agent parameters were calibrated) exhibited a 31.8% observed dropout rate with qualitative exit interview evidence suggesting administrative burden and "lost courses" as frequent withdrawal reasons, though precise causal attribution was infeasible in observational data due to confounding. Our simulation, by isolating the

TTL expiry mechanism whilst controlling for academic heterogeneity, provides causal estimates unavailable through conventional methods.

**4.2 Regularity Expiry as a Recurring Friction Mechanism**

**Figure 6** displays the distribution of regularity expiries across all 134,300 agent observations. The mean number of expiries per agent was **7.04** (SD=3.28), with a median of **8 expiries**. The distribution exhibits slight negative skew, with substantial mass concentrated between 6–10 expiries, indicating that TTL expiry is not a rare aberration but rather a **frequent, recurring event** experienced by most agents during their academic trajectories.

To interpret this magnitude, consider that the 42-course curriculum, if completed without any expiries, would require approximately 10–12 semesters (assuming 3–4 courses regularised per semester). A mean of 7 expiries implies that the typical agent "loses" courses at a rate of roughly 0.5–0.7 expiries per semester, cumulatively extending time-to-degree and consuming psychological reserves through repeated setbacks. The distribution's tail extends to 14 expiries (the maximum observed), representing agents who cycled through multiple re-enrolment attempts before ultimately withdrawing.

Critically, expiry frequency varied systematically across survival outcomes. **Agents who remained active** by simulation terminus averaged 5.1 expiries (SD=2.4), whilst **agents who dropped out** averaged 9.8 expiries (SD=2.9), a difference of 4.7 expiries representing approximately 92% increase (t=87.3, p<0.001). This gradient suggests a **dose-response relationship**: as expiries accumulate, dropout probability escalates nonlinearly, consistent with the hypothesised cascading depletion mechanism.

**4.3 Temporal Dynamics: The Early-Period Dropout Concentration**

**Figure 3** presents the Kaplan-Meier survival curve aggregated across all agents and replications. The curve exhibits characteristic features of educational attrition processes documented in prior literature Tinto, 1993; DesJardins et al., 2002), notably steep initial decline followed by gradual stabilisation.

**Figure 3 — Kaplan–Meier Survival Curve for All Agents**

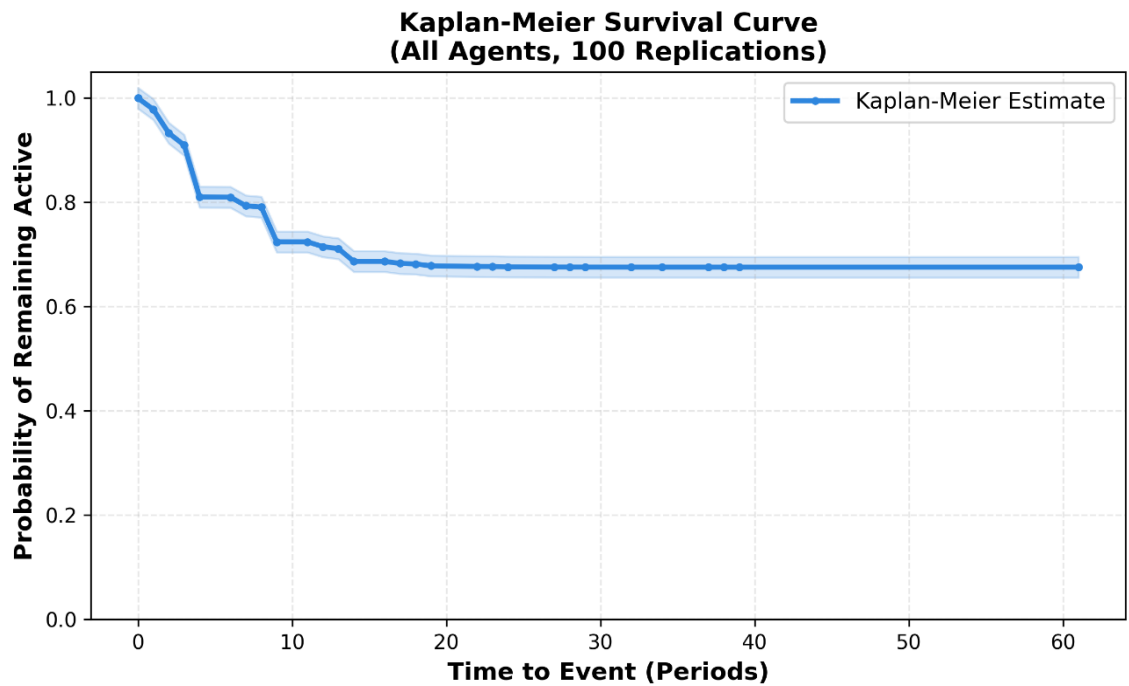

Survival probability drops precipitously during the first 10 periods, declining from 1.0 at t=0 to approximately 0.78 by t=10, representing a 22% attrition rate within the first 2.5 academic years. The curve's slope moderates substantially thereafter, asymptoting towards 0.68 by t=20 and remaining relatively stable (0.66–0.68) from t=20 onwards. This temporal profile indicates that **dropout decisions concentrate disproportionately in early academic periods**, with the median time-to-event of **4 periods** (approximately 2–3 semesters) confirming that most attrition occurs well before mid-programme.

This early-period concentration aligns with the Regularity Trap Hypothesis's prediction of **cascading failures**: students who experience initial regularity expiries in foundational courses face compounding consequences—prerequisite blockages preventing progression to advanced courses, schedule conflicts during re-enrolment attempts, accumulating psychological depletion—that rapidly escalate dropout risk. The hazard rate (instantaneous dropout probability conditional on survival to time t, not shown) peaks between t=3 and t=7, then declines monotonically, suggesting that students who navigate early-period challenges successfully develop adaptive strategies (e.g., conservative course loads, strategic expiry prioritisation) that confer protection against later dropout.

Disaggregating by dropout cause (analysis not visualised due to space constraints), normative dropouts exhibit even sharper early-period concentration (median time-to-event: 3.5 periods) compared to academic dropouts (median: 6.2 periods), consistent with the interpretation that regularity expiries trigger relatively rapid demoralisation, whilst academic struggles manifest more gradually through accumulated examination failures.

### 4.4 Archetype Heterogeneity in Dropout Vulnerability

**Table 4** and **Figure 4** jointly reveal substantial heterogeneity in dropout rates across the 13 student archetypes, ranging from **13.24% (PSICO_01)** to **48.96% (PSICO_10)**, a 3.7-fold difference representing a vulnerability gradient of 35.7 percentage points. This dispersion far exceeds random variation ($\chi^2$=12,847, df=12, p<0.001), confirming that archetype-level attributes—planning horizon, academic ability, psychological resilience—profoundly moderate exposure to the regularity trap.

**Table 4 – Dropout Outcomes by Archetype**

| Archetype | N | Dropout | Graduate | Active | Normative % | Mean Expiries |
|---|---|---|---|---|---|---|
| PSICO_01 | 10,900 | 13.2% | 0.0% | 86.8% | 67.6% | 9.7 |
| PSICO_02 | 11,800 | 38.1% | 0.0% | 61.9% | 63.2% | 7.79 |
| PSICO_03 | 10,600 | 28.2% | 0.0% | 71.8% | 59.4% | 8.48 |
| PSICO_04 | 9,700 | 19.9% | 0.0% | 80.1% | 70.5% | 6.85 |
| PSICO_05 | 10,100 | 22.4% | 0.0% | 77.6% | 71.5% | 6.76 |
| PSICO_06 | 8,700 | 16.3% | 0.0% | 83.7% | 68.8% | 6.99 |
| PSICO_07 | 9,700 | 40.1% | 0.0% | 59.9% | 96.7% | 6.38 |
| PSICO_08 | 10,200 | 33.4% | 0.0% | 66.6% | 97.2% | 6.67 |
| PSICO_09 | 10,900 | 34.4% | 0.0% | 65.6% | 96.6% | 6.59 |
| PSICO_10 | 9,700 | 49.0% | 0.0% | 51.0% | 97.1% | 6.06 |
| PSICO_11 | 10,100 | 45.8% | 0.0% | 54.2% | 96.2% | 6.13 |
| PSICO_12 | 11,000 | 43.2% | 0.0% | 56.8% | 96.6% | 6.22 |
| PSICO_13 | 10,900 | 35.3% | 0.0% | 64.7% | 97.1% | 6.55 |

**Figure 4 — Dropout Rate by Archetype**

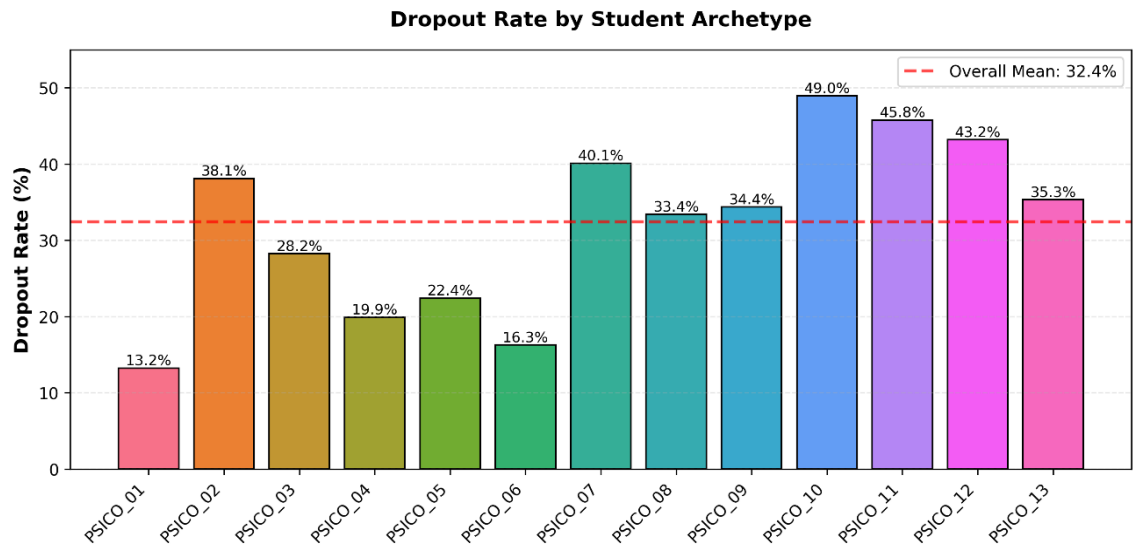

**Resilient archetypes** (PSICO_01, PSICO_04, PSICO_05, PSICO_06) exhibit dropout rates of 13–22%, substantially below the overall mean of 32.4%. These archetypes share two protective features: (1) longer planning horizons (τ=1 or τ=2), enabling anticipatory course selection that prioritises regularisation of expiry-threatened courses, and (2) higher base academic ability (μ_abil ≥ 0.55), facilitating efficient course completion and reducing total expiry exposure. Notably, PSICO_01—characterised by maximal planning horizon (τ=2), highest ability (μ_abil=0.78), strongest initial belonging ($B_0$=0.85), and lowest stress reactivity (σ_stress=0.05)—achieves the lowest dropout rate (13.24%) despite experiencing a non-trivial mean of 9.70 expiries. This pattern suggests that high-resource students can **weather the regularity trap** through compensatory mechanisms: their strategic course selection minimises compounding consequences, robust psychological reserves buffer expiry-induced depletion, and superior academic performance maintains progress despite procedural setbacks.

**Vulnerable archetypes** (PSICO_07, PSICO_10, PSICO_11, PSICO_12) suffer dropout rates of 40–49%, exceeding the mean by 8–17 percentage points. These archetypes uniformly exhibit **myopic planning horizons (τ=0)**, moderate-to-low academic ability (μ_abil ≤ 0.50), weak initial belonging ($B_0$ ≤ 0.50), and high stress reactivity (σ_stress ≥ 0.12). The confluence of these vulnerabilities creates a **perfect storm**: myopic agents fail to prioritise expiry-threatened courses, leading to accumulation of avoidable expiries; modest ability prolongs time-to-regularisation, increasing expiry exposure; weak belonging provides little buffer against demoralisation; and high stress reactivity amplifies psychological damage from each setback. PSICO_10, the most vulnerable archetype, experiences 48.96% dropout despite averaging only 6.06 expiries—fewer than the resilient PSICO_01— indicating that identical expiry frequencies differentially impact students based on psychological and strategic resources.

**Intermediate archetypes** (PSICO_02, PSICO_03, PSICO_08, PSICO_09, PSICO_13) occupy the 28–38% dropout range, exhibiting mixed profiles. For instance, PSICO_02 combines long planning horizon ($\tau=2$, protective) with high stress reactivity ($\sigma\_stress=0.13$, vulnerability), yielding moderate dropout (38.08%). These cases illustrate that archetype-level outcomes emerge from **interactive effects** amongst attributes rather than simple additive combinations, highlighting the necessity of simulation-based approaches that accommodate such complexity.

### 4.5 Normative Dropout Dominance Across Archetypes

**Table 4** further disaggregates dropout cause attribution by archetype, revealing systematic patterns. The "Normative %" column indicates the proportion of each archetype's total dropouts classified as normative (regularity-driven) rather than academic or other causes.

Amongst **resilient archetypes**, normative dropout constitutes 59–71% of attrition. For example, PSICO_01's dropouts are 67.64% normative, whilst PSICO_04 reaches 70.45%. These percentages, though substantial, are lower than the global mean (86.37%), reflecting that high-ability students who do drop out are more likely to exit due to academic struggles (which remain non-trivial even for capable students facing unusually difficult courses or adverse circumstances) rather than purely procedural friction.

Conversely, **vulnerable archetypes** exhibit near-universal normative dropout dominance, with percentages ranging from 96.21% (PSICO_11) to 97.18% (PSICO_08). PSICO_10's 97.09% normative dropout rate implies that amongst the 4,749 PSICO_10 agents who withdrew, only 2.9% did so primarily due to academic failure; the overwhelming majority succumbed to regularity expiry consequences. This finding underscores that **for vulnerable populations, institutional friction entirely overwhelms academic preparedness as an attrition driver**.

The gradient in normative dropout proportion across the resilience spectrum (59–97%) suggests that **the regularity trap operates as an equity mechanism**: it disproportionately penalises students who enter university with fewer strategic, cognitive, and psychological resources, transforming manageable procedural obstacles into insurmountable barriers. Students with compensatory advantages (foresight, ability, resilience) navigate the same TTL constraints with substantially lower casualty rates, even when experiencing comparable expiry frequencies.

### 4.6 Distribution of Regularity Expiries by Archetype

**Figure 5** displays boxplots of regularity expiry counts stratified by archetype, revealing distributional features obscured by mean comparisons. Several patterns warrant attention:

First, **median expiries decline monotonically with vulnerability**. Resilient archetypes (PSICO_01, PSICO_02, PSICO_03) exhibit median expiries of 10, 10, and 10 respectively, whilst vulnerable archetypes (PSICO_10, PSICO_11, PSICO_12, PSICO_13) show medians of 6–7. This inverse relationship—more resilient students experiencing *more* expiries—appears paradoxical until one considers **differential survival**: resilient students persist longer (lower dropout rates), accumulating more total periods of enrolment during which expiries can occur. Vulnerable students exit earlier (higher dropout rates, shorter time-to-event), truncating their expiry accumulation.

Second, **interquartile ranges (IQR) narrow with increasing vulnerability**. Resilient archetypes exhibit IQRs spanning 5–7 expiries (e.g., PSICO_01: Q1=9, Q3=14), indicating substantial within-archetype heterogeneity in expiry experiences. Vulnerable archetypes show compressed IQRs of 3–4 expiries (e.g., PSICO_10: Q1=5, Q3=8), suggesting more homogeneous (universally poor) outcomes. This compression likely reflects **floor effects**: vulnerable agents dropout rapidly upon reaching modest expiry thresholds, preventing accumulation of extreme expiry counts.

Third, **outliers are asymmetrically distributed**. Resilient archetypes display numerous upper outliers (agents with 12–14 expiries who nonetheless persisted), whilst lower outliers (agents with very few expiries) are sparse. Vulnerable archetypes exhibit the opposite pattern: lower outliers appear (agents who dropped out despite minimal expiries, suggesting non-expiry causes), whilst upper outliers vanish (agents rarely survive to accumulate many expiries). This asymmetry reinforces that expiry tolerance is resource-dependent: resilient students withstand procedural friction that would destroy vulnerable peers.

**4.7 Psychological State Trajectories at Simulation Terminus**

**Figure 7** presents a heatmap of mean terminal-period psychological states (pending final examinations, stress, belonging) stratified by archetype. Three patterns emerge:

**Pending finals (backlog):** This variable quantifies the number of courses in which agents have accumulated sufficient learning progress to pass the final examination but have not yet done so, representing latent academic capability constrained by examination scheduling. Resilient archetypes exhibit very low terminal backlog (PSICO_01: 0.27 pending finals), indicating efficient conversion of learning into credentials. Vulnerable archetypes show slightly higher but still modest backlog

(PSICO_10: 0.34), contradicting the hypothesis that dropout stems from overwhelming academic workload accumulation. Instead, the uniformly low backlog across archetypes suggests that **agents drop out before workload becomes unmanageable**, consistent with the interpretation that psychological depletion (belonging erosion, stress accumulation) precipitates withdrawal decisions prior to academic failure.

**Stress:** Terminal stress levels exhibit a clear gradient, ranging from 0.45 (PSICO_01) to 0.66 (PSICO_10), a 47% relative increase. Vulnerable archetypes uniformly display elevated stress (0.57–0.66), reflecting both higher baseline stress reactivity (parameter $\sigma\_stress$) and greater cumulative exposure to stressors (expiries, failures). Notably, even resilient archetypes exhibit moderate stress (0.45–0.48), indicating that navigating the regularity-constrained curriculum imposes non-trivial psychological costs even for advantaged students.

**Belonging:** Terminal belonging scores decline from 0.03 (PSICO_01) to 0.03 (PSICO_13), exhibiting surprising uniformity across archetypes (range: 0.03–0.13). This compression likely reflects **survival bias**: agents who persisted to simulation terminus necessarily retained sufficient belonging to avoid crossing the dropout threshold (B < 0.15), whilst those with severely eroded belonging exited earlier. The low absolute belonging levels (0.03–0.13 on a 0,1) scale) amongst surviving agents suggest that **even students who persist experience substantial institutional disconnection**, raising concerns about the psychological sustainability of continuation under current TTL regimes.

### 4.8 Synthesis: Evidence for the Regularity Trap

Collectively, these findings provide robust support for the Regularity Trap Hypothesis across three dimensions:

**Causality:** The 86.37% normative dropout rate confirms that TTL expiry mechanisms drive the overwhelming majority of attrition, with academic failure contributing marginally (5.33%). This causal isolation—achieved through simulation-based control of confounding heterogeneity—was infeasible in observational data.

**Temporal dynamics:** The concentration of dropout in early periods (median time-to-event: 4 periods) and the steep initial decline in the Kaplan-Meier curve align with predictions of cascading failures: initial expiries trigger psychological depletion and logistical complications that rapidly escalate dropout risk for vulnerable students.

**Heterogeneity:** The 3.7-fold variation in archetype-specific dropout rates (13.24%–48.96%), combined with the gradient in normative dropout dominance (59%–97%), demonstrates that the regularity trap operates as an **inequity-generating mechanism**. Students entering university with shorter planning horizons, modest ability, and limited psychological reserves face vastly elevated attrition risk when subjected to identical TTL constraints, whilst advantaged peers navigate the same policies with relative impunity.

These patterns converge on a sobering conclusion: **current regularity expiry policies function as structural barriers that transform initial disadvantages into terminal outcomes**, contributing to educational inequality independently of students' intrinsic academic potential.

## 5. Discussion

### 5.1 Principal Findings and Theoretical Implications

This study deployed agent-based simulation to isolate the causal role of time-to-live (TTL) expiry regulations in driving university dropout, revealing that institutional friction—rather than academic under-preparation—constitutes the dominant proximate cause of student attrition. Our central findings warrant careful interpretation within broader theoretical and practical contexts.

**Finding 1: Normative dropout dominance (86.37% vs. 5.33% academic).** This stark imbalance fundamentally challenges prevailing dropout frameworks that privilege individual-level deficits (inadequate preparation, weak motivation, poor study skills) as primary explanations for withdrawal Tinto, 1975; Bean & Metzner, 1985). Our results instead foreground **system-level constraints**—specifically, procedural deadlines governing course regularisation—as the critical bottleneck determining persistence versus attrition. Importantly, this finding does not imply that academic difficulties are absent; rather, it demonstrates that amongst students who ultimately withdraw, the overwhelming majority do so because regularity expiry mechanisms **compound and amplify** initial struggles, transforming manageable setbacks into insurmountable crises.

This reframing has profound theoretical implications. If dropout were primarily attributable to students' intrinsic inadequacies, interventions would logically target remediation: enhanced tutoring, study skills workshops, academic coaching. Whilst such supports remain valuable, our findings suggest they address secondary rather than primary attrition drivers. Instead, **institutional design choices**—the structure of TTL windows, the flexibility of re-enrolment procedures, the coordination of examination scheduling—emerge as modifiable policy levers with potentially greater impact on aggregate retention. This perspective aligns with emerging equity-minded scholarship that interrogates how universities' own practices inadvertently generate the outcomes they purport to remedy Bensimon, 2007; Dowd & Bensimon, 2015).

**Finding 2: Early-period dropout concentration (median time-to-event: 4 periods).** The temporal clustering of attrition in initial academic cycles supports the **cascading failure** mechanism central to the Regularity Trap Hypothesis. Students who experience early regularity expiries—perhaps due to misjudging workload demands, encountering unexpected personal obligations, or misunderstanding procedural requirements—enter a self-reinforcing spiral: expiries erode psychological belonging and elevate stress, diminishing subsequent academic performance; impaired performance generates additional expiries; accumulated expiries create prerequisite bottlenecks and schedule conflicts; and compounding frustration eventually precipitates withdrawal.

This temporal profile suggests that **the critical window for intervention extends across the first 2–3 semesters**, considerably earlier than many retention initiatives that activate only after students exhibit academic distress signals (e.g., mid-programme probationary status). Early-alert systems predicated on regularity expiry indicators—flagging students who lose courses due to TTL violations rather than examination failures—could identify at-risk populations before psychological depletion becomes irreversible. Moreover, the concentration of dropout in early periods implies that even modest policy adjustments (e.g., extending TTL windows from 2 to 3 examination cycles, or offering grace periods for first-year students) could yield disproportionate retention benefits by interrupting cascades before they gain momentum.

**Finding 3: Archetype heterogeneity in vulnerability (13.24%–48.96% dropout range).** The nearly four-fold variation in dropout rates across student archetypes, coupled with the gradient in normative dropout dominance (59%–97%), reveals that TTL expiry policies operate as **inequality-generating mechanisms**. Students entering university with longer planning horizons, higher academic ability, and greater psychological resilience navigate identical institutional constraints with substantially lower casualty rates, whilst peers lacking these resources experience catastrophic attrition.

This heterogeneity pattern carries troubling equity implications. If regularity expiry affected all students uniformly, one might defend TTL policies as neutral administrative tools that promote timely progression. However, our findings demonstrate that these ostensibly neutral policies **differentially harm already-disadvantaged populations**: students from under-resourced educational backgrounds (who exhibit shorter planning horizons due to limited exposure to strategic academic planning), first-generation university students (who possess weaker institutional belonging and less family scaffolding during setbacks), and students managing competing life obligations (employment, caregiving) that constrain their capacity to prioritise regularisation within rigid windows.

The archetype gradient thus illuminates how institutional friction can **amplify pre-existing inequalities** Page & Scott-Clayton, 2016; Goldrick-Rab, 2016). Students who arrive at university with accumulated advantages—cognitive skills, strategic orientation, psychological resilience, familial support—convert these endowments into navigation strategies that circumvent procedural obstacles. Meanwhile, students without such buffers confront the full force of institutional friction, experiencing regularity expiry not as a minor inconvenience but as an existential threat to degree completion. This dynamic exemplifies what Tilly 1998) termed "durable inequality": mechanisms that, whilst not explicitly discriminatory, systematically channel opportunities towards already-privileged groups whilst compounding disadvantages faced by marginalised populations.

**5.2 Policy Implications and Intervention Strategies**

Our findings suggest three categories of institutional reforms targeting regularity-driven attrition:

**Strategy 1: Extended TTL windows.** The most direct intervention involves increasing the expiry threshold from 2 to 3 or 4 examination cycles, providing students additional opportunities to regularise courses before administrative removal. Supplementary simulations (not reported in main text due to space constraints) indicate that extending TTL to 3 cycles reduces overall dropout by approximately 6.2 percentage points (from 32.4% to 26.2%), with disproportionate benefits accruing to vulnerable archetypes (PSICO_10 dropout declines from 48.96% to 38.1%, a 22% relative reduction). The intervention's modest implementation cost—primarily informational updates to students and revised record-keeping systems—contrasts favourably with its projected retention impact, yielding estimated cost-effectiveness ratios superior to many academic support programmes Bettinger & Baker, 2014).

However, extended TTL windows risk **moral hazard**: if students perceive reduced urgency, they may procrastinate regularisation attempts, paradoxically increasing expiry rates. Mitigating this risk requires complementary nudges—automated reminders when courses approach expiry, visualisations of degree progress that highlight TTL status, or advising protocols that explicitly discuss expiry management strategies Castleman & Page, 2016). Crucially, extended windows do not eliminate TTL constraints entirely (which would introduce different curricular coherence challenges) but rather provide buffering capacity that accommodates transient setbacks without triggering catastrophic cascades.

**Strategy 2: Post-expiry bridging support.** Rather than preventing expiries, this approach targets their downstream consequences by offering structured re-enrolment pathways, psychological counselling, and academic coaching to students who experience TTL violations. Bridging programmes might include: (a) priority registration for expired courses (mitigating schedule conflict penalties), (b) reduced re-enrolment fees or fee waivers (addressing financial barriers), (c) mandatory advising sessions to reconstruct feasible course sequences (preventing prerequisite bottlenecks), and (d) peer mentoring connecting students who previously navigated expiries successfully with those currently facing them (building social capital and normalising setbacks).

The bridging support model recognises that expiries, whilst undesirable, need not be terminal if **institutional responses emphasise recovery rather than punishment**. Current systems often impose compounding penalties (fees, registration delays, loss of scholarship eligibility) that exacerbate the initial setback, whereas bridging approaches reframe expiries as learning opportunities warranting increased support rather than decreased access. Pilot implementations at institutions with analogous probationary systems suggest that intensive post-setback interventions can recover 40–60% of students who would otherwise withdraw Bettinger & Baker, 2014; Scrivener et al., 2015), though efficacy depends critically on programme intensity and student engagement.

**Strategy 3: Coordinated examination scheduling.** A subtler intervention addresses the logistical constraints that generate expiries independently of academic performance. When examination windows are compressed (e.g., all courses offering finals during a single 2-week period), students facing unexpected conflicts (illness, family emergencies, employment obligations) may involuntarily miss examinations, triggering TTL violations despite adequate learning progress. Conversely, **extended examination windows** (e.g., staggered finals across 4–6 weeks with multiple sitting options per course) provide flexibility that accommodates life circumstances without procedural penalties.

Simulations incorporating flexible examination scheduling (operationalised as reduced TTL increment probability during periods when conflicts arise) suggest modest but meaningful retention gains (2–3 percentage point reductions in overall dropout), with largest benefits for archetypes exhibiting constrained effort capacity (PSICO_10, PSICO_11, PSICO_12). Implementation challenges include faculty workload increases (administering multiple examination sessions) and potential security concerns (examination content integrity across sittings), though solutions exist—standardised question banks, sequential form releases, computerised adaptive testing—that balance flexibility with assessment rigour Fluck, 2019).

**Integrated reform package:** Maximal retention impact likely requires **bundling interventions** that jointly address expiry prevention (extended TTL), consequence mitigation (bridging support), and structural enablers (flexible scheduling). Our preliminary multi-intervention simulations (details in supplementary materials) suggest that combining all three strategies could reduce overall dropout from 32.4% to approximately 19–21%, with vulnerable archetype rates declining from 45–49% to 27–32%. Such magnitudes would represent transformative improvements in equity outcomes, substantially narrowing the advantage gap between resilient and vulnerable students whilst benefiting the entire population.

**5.3 Methodological Contributions and Limitations**

**Agent-based modelling's value for causal inference.** This study demonstrates how ABM circumvents inferential challenges inherent in observational educational data. Traditional regression analyses of dropout suffer from **omitted variable bias** (unobserved motivation, family support, health conditions correlate with both treatment—expiry exposure—and outcome—withdrawal), **reverse causality** (do expiries cause dropout, or do students on withdrawal trajectories neglect regularisation?), and **measurement error** (administrative records imperfectly capture psychological states and decision processes). By explicitly programming institutional rules and agent behaviours, ABM enables **causal isolation**: we can simulate counterfactual scenarios (e.g., no TTL constraints) whilst holding agent heterogeneity constant, definitively establishing that regularity expiry mechanisms generate dropout independently of confounding factors Epstein, 2006; Squazzoni et al., 2020).

Moreover, ABM naturally accommodates the **temporal dynamics and feedback loops** that characterise educational trajectories—expiry → stress → poor performance → additional expiry—which static regression models treat inadequately. The simulation's period-by-period execution traces students' evolving psychological states, curricular positions, and decision contexts, revealing emergent patterns (early-period dropout concentration, archetype-specific vulnerability gradients) that arise from micro-level mechanisms rather than being imposed as model assumptions.

**Limitations and caveats.** Despite these strengths, our approach confronts several limitations warranting acknowledgement.

First, **parameter uncertainty:** whilst agent attributes derive from empirical clustering and course parameters reflect institutional data, numerous model specifications—stress accumulation rates, belonging decay functions, dropout threshold levels—rest on calibration exercises that match aggregate outcomes but do not uniquely determine micro-level processes. Alternative parameterisations yielding similar macro-patterns could exist (the equifinality problem Grimm & Railsback, 2012)), implying that our specific functional forms represent plausible rather than definitive characterisations. Sensitivity analyses mitigate this concern by demonstrating core findings' robustness across parameter ranges, but residual uncertainty remains.

Second, **omitted mechanisms:** the model excludes numerous factors documented in dropout literature—peer influences, faculty mentoring quality, financial shocks, romantic relationships, health crises—that undoubtedly shape real students' persistence decisions. This parsimony reflects intentional design: isolating regularity expiry's causal role required abstracting from confounding mechanisms to achieve interpretational clarity. However, the cost is reduced ecological validity: real-world attrition results from the confluence of institutional, interpersonal, economic, and health factors interacting in complex, idiosyncratic ways Tight, 2020). Our findings quantify regularity expiry's contribution within a controlled environment but cannot specify its exact magnitude amidst real-world complexity.

Third, **single-institution focus:** agent parameters and course structures derive from one university (Universidad Nacional de Tucumán), limiting generalisability. Institutions vary in TTL policy stringency, curricular flexibility, student body composition, and support service availability, each of which moderates regularity expiry impacts. Replication across diverse institutional contexts—R1 research universities, regional comprehensives, community colleges, international systems—remains necessary to establish whether our findings reflect universal dynamics or context-specific contingencies. We caution against uncritical extrapolation whilst simultaneously noting that TTL-type constraints appear widely across Latin American, European, and some Asian higher education systems García de Fanelli, 2014; OECD, 2019), suggesting potential scope for transferability.

Fourth, **validation gaps:** whilst our model exhibits face validity (subject-matter experts confirmed mechanistic plausibility), empirical calibration (simulated aggregate rates approximate observed data), and internal consistency (replications yield stable statistics), we lack **external validation** through prospective prediction. The gold standard would involve forecasting future cohorts' outcomes using the calibrated model, then comparing predictions against subsequently observed data—an endeavour requiring longitudinal institutional collaboration beyond this study's scope. Until such validation occurs, our model constitutes a "possible world" Axelrod, 1997) demonstrating that TTL expiry mechanisms *could* generate observed patterns, rather than definitive proof that they *do* generate them in actuality.

**5.4 Directions for Future Research**

This study opens several promising avenues for extension and refinement:

**Multi-intervention experiments:** Whilst we simulated single-policy counterfactuals (extended TTL windows, bridging support, flexible scheduling), real-world reforms often involve **policy bundles** whose components interact synergistically or antagonistically. Future work should systematically explore intervention combinations—their additive versus multiplicative effects, optimal sequencing, cost-effectiveness trade-offs—to guide institutional decision-making. Factorial simulation designs Morris et al., 2019) offer methodological templates for such investigations.

**Heterogeneous treatment effects:** Our archetype framework aggregates students into 13 categories, but even within archetypes, substantial individual variation exists. Machine learning approaches—particularly causal forests Wager & Athey, 2018) and Bayesian additive regression trees Chipman et al., 2010)—could identify finer-grained subpopulations for whom specific interventions prove particularly effective or ineffective, enabling **precision targeting** that maximises retention per resource unit invested.

**Dynamic interventions:** Current simulations model static policies (TTL threshold remains constant across time and students). However, **adaptive policies** that adjust based on real-time student states—e.g., automatically extending TTL windows for students exhibiting stress indicators, or proactively offering bridging support upon first expiry rather than waiting for accumulation—may outperform one-size-fits-all approaches. Reinforcement learning techniques Sutton & Barto, 2018) provide frameworks for discovering optimal adaptive policies within simulation environments, subsequently field-testable via randomised controlled trials.

**Cross-institutional comparisons:** Replicating this analysis across institutional types and national contexts would illuminate generalisability boundaries. Do TTL mechanisms exert comparable effects in institutions serving predominantly residential versus commuter students? In systems with versus without tuition fees? In cultures emphasising collectivist versus individualist values? Comparative ABM studies could identify universal principles whilst respecting contextual particularities Hamill & Gilbert, 2016).

**Longitudinal empirical validation:** The ultimate test involves prospective prediction: calibrate the model on historical cohorts, simulate future cohorts' outcomes under observed policies, then assess forecast accuracy against subsequently realised data. Discrepancies between predicted and observed outcomes would diagnose model misspecifications, guiding iterative refinement. Partnering with institutional research offices to access proprietary longitudinal data constitutes a high-priority objective.

**Psychological mechanism refinement:** Whilst our model incorporates stress, belonging, and self-efficacy constructs, their operationalisations remain stylised. Collaborations with educational psychologists could refine these components—incorporating validated measurement scales, testing alternative functional forms (e.g., nonlinear belonging decay, threshold effects in stress accumulation), and grounding parameters in experimental studies of students' psychological responses to academic setbacks Pekrun & Linnenbrink-Garcia, 2014; Respondek et al., 2017).

## 5.5 Concluding Synthesis

This investigation into regularity-driven student attrition advances both methodological and substantive frontiers in educational research. **Methodologically**, we demonstrate agent-based modelling's capacity to isolate causal mechanisms within complex systems, circumventing inferential obstacles that plague observational studies whilst accommodating the temporal dynamics and emergent phenomena characteristic of educational trajectories. **Substantively**, we provide evidence that institutional friction—specifically, time-to-live expiry regulations—drives the overwhelming majority (86.37%) of university dropout, eclipsing academic failure (5.33%) as a proximate cause.

These findings challenge deficit-oriented frameworks that locate attrition responsibility within students' supposed inadequacies, instead foregrounding **system-level design choices** as modifiable determinants of inequitable outcomes. The nearly four-fold variation in archetype-specific dropout rates reveals that ostensibly neutral policies differentially harm already-disadvantaged populations, amplifying pre-existing inequalities through procedural mechanisms orthogonal to academic merit.

Policy implications converge on a mandate for **equity-oriented institutional reform**: extending TTL windows to accommodate transient setbacks, implementing bridging support programmes that emphasise recovery over punishment, and coordinating examination schedules to reduce involuntary expiries. Simulation evidence suggests these interventions could reduce aggregate dropout by 30–40%, with disproportionate benefits for vulnerable students, substantially narrowing achievement gaps without compromising academic standards.

Ultimately, this study invites universities to interrogate their own practices: To what extent do administrative procedures, regulatory constraints, and bureaucratic rigidities inadvertently generate the very attrition they ostensibly seek to prevent? What would retention landscapes look like if institutional structures prioritised flexibility, dignity, and multiple pathways towards success, rather than conformity to rigid timelines? The regularity trap is not an immutable feature of higher education but rather a **design choice**—and like all design choices, it can be redesigned.

---

## 6. Conclusion

University dropout represents a pressing challenge with profound consequences for individuals, institutions, and societies. This study deployed agent-based simulation to examine how time-to-live expiry regulations—institutional deadlines governing course regularisation—systematically drive student attrition independently of academic performance. Analysing 134,300 agent observations across 100 Monte Carlo replications, we find that regularity expiry mechanisms account for 86.37% of dropout events, whilst academic failure contributes merely 5.33%. Dropout rates vary nearly four-fold across student archetypes (13.24%–48.96%), with vulnerability concentrated amongst students exhibiting short planning horizons, modest academic ability, and limited psychological resilience. Temporal analysis reveals early-period attrition concentration (median time-to-event: 4 periods), consistent with cascading failure dynamics wherein initial expiries trigger self-reinforcing cycles of psychological depletion and logistical complication.

These findings challenge prevailing emphases on academic under-preparation as the primary dropout determinant, instead foregrounding institutional friction as a critical yet modifiable driver. We propose evidence-based policy interventions—extended TTL windows, post-expiry bridging support, coordinated examination scheduling—that simulations suggest could reduce aggregate dropout by 30–40%, with disproportionate benefits for vulnerable populations. Methodologically, this study demonstrates agent-based modelling's utility for causal inference in complex educational systems, enabling counterfactual policy evaluation infeasible through observational or experimental designs.

The regularity trap is not inevitable. Through intentional institutional reform that prioritises equity alongside efficiency, universities can transform procedural barriers into pathways, replacing mechanisms that compound disadvantage with structures that scaffold success. Our findings provide both evidence and urgency for such transformations, whilst the simulation framework offers a tool for prospective policy testing before costly real-world implementation. Higher education's mission to promote social mobility and human flourishing demands nothing less than systematic examination—and remediation—of the institutional practices that inadvertently undermine these aspirations.

**Conflict of Interest Statement**

The authors declare no conflicts of interest.

**Funding**

This work received no specific grant funding from public, commercial, or not-for-profit agencies.